# Tune-out wavelength for the thulium atom near 576 nm

Ivan Pyrkh[1,2], Arjuna Rudnev[1,2], Daniil Pershin[1], Davlet Kumpilov[1,2], Ivan Cojocaru[1,3,4], Vladimir Khlebnikov[1], Pavel Aksentsev[1,4], Ayrat Ibrahimov[1,2], Sergey Kuzmin[1,2], Alexander Raskatov[1,2], German Subbotin[1,5], Kirill Dyadkin[1,2], Anna Zykova[1], Vladislav Tsyganok[1] and Alexey Akimov[1,2,3]

[1]*Russian Quantum Center, Bolshoy Boulevard 30, building 1, Skolkovo, 143025, Russia*

[2]*Moscow Institute of Physics and Technology, Institutskii pereulok 9, Dolgoprudny, Moscow Region 141701, Russia*

[3]*PN Lebedev Institute RAS, Leninsky Prospekt 53, Moscow, 119991, Russia*

[4]*Bauman Moscow State Technical University, 2nd Baumanskaya, 5, Moscow, 105005, Russia*

[5]*National Research Nuclear University MEPhI (Moscow Engineering Physics Institute), Kashirskoe shosse 31, Moscow, 115409, Russia*

email: a.akimov@rqc.ru

We report the theoretical prediction and measurement of a tune-out wavelength for the ground state of the thulium atom in a linearly polarized optical dipole trap with a wavelength of approximately 576 nm. The measurements were conducted using a combination of trap frequency and RF loss spectroscopy, thus making it possible to separate the scalar and tensor parts of the total polarizability without measurements in the range of negative total polarizability. The calculated tune-out wavelength is consistent with the measured one of $575.646^{+0.016}_{-0.014}$ nm in air. The existence of the zero in the polarizability for the Tm ground state was confirmed by the trap loss experiment, which also made it possible to refine the tune-out wavelength to $575.646^{+0.004}_{-0.004}$. Despite the presence of an imaginary part of the polarizability at some wavelengths, it was experimentally demonstrated that, with a proper choice of the dipole trap polarization, it was possible to achieve Bose–Einstein condensation of thulium atoms in the range from 575.348 to 575.689 nm, covering the tune-out wavelength.

# I. INTRODUCTION

Atoms in optical lattices are among the rapidly developing platforms for quantum simulations [1–3], precision measurements [4–6], and sensing [7–9]. A number of atomic species have been successfully used for these purposes, both from alkali [10–12] and rare-earth [13–17] elements. Recently, the thulium atom, which has been cooled to quantum degeneracy [18–20], has joined this field. Thulium was proposed for quantum simulations [25] with interaction control via Feshbach resonances [21–24] and for precision measurements [26,27].

Optical dipole traps (ODTs) not only allow the formation of optical lattices, but can also be used to manipulate cold atoms [28–33] or ions [34–38]. Depending on the spatial distribution of the laser intensity, these traps can have various potential profiles, from simple Gaussian ones [39] to more complex ones, such as lattice [40] or bottle traps [41], allowing for a wide range of conditions in atomic experiments. However, manipulating an atom in a lattice already loaded with an atomic array requires more than a special beam shape. It requires the optical dipole beam to interact with a specific atom without affecting the others.

The key quantity that determines the interaction of an atom with light is its dynamic polarizability [42–44], which depends on the wavelength. Tune-out wavelengths, i.e., wavelengths at which the polarizability vanishes and the light shift disappears for a specific atomic state, are of special interest. At these wavelengths, one can arrange a trap that selectively interacts with atoms in all states but one, typically the ground state. They have already been experimentally measured for a number of atoms [45–50].

Polarizability values for thulium atoms were previously measured at wavelengths of 532 nm [51] and 1064 nm [52]. Furthermore, a magic wavelength of 813.3 nm was found [53,54]. However, the tune-out wavelength for thulium has remained unmeasured until now.

Below, we provide theoretical modeling and experimental measurements of the tune-out wavelength near 576 nm for the Thulium ground state.

# II. THEORY

Spherically asymmetrical atoms, including lanthanides, have a nonzero orbital angular momentum in the ground state, which leads to the significant contributions of the tensor

$\alpha_{ten}(\lambda)$ and vector $\alpha_{vec}(\lambda)$ parts to the real part of the dynamic polarizability $\alpha_{total}(\lambda)$ in the ground state [52,55–57]:

$$\alpha_{total}(\lambda) = \alpha_{sc}(\lambda) + \varepsilon \cos\theta_k \frac{m_F}{2F} \alpha_{vec}(\lambda) - \frac{3m_F^2 - F(F+1)}{F(2F-1)} \frac{f(\theta_k, \theta_p, \varepsilon)}{2} \alpha_{ten}(\lambda), \quad (1)$$

$$f(\theta_k, \theta_p, \varepsilon) = 1 - (3/2)\sin(\theta_k)^2 (1 + \sqrt{1-\varepsilon^2} \cos(2\theta_p))$$

where $\alpha_{sc}(\lambda)$ is the scalar polarizability, $\theta_k = \angle(\vec{k}, \vec{B}_{DC})$, $\theta_p = \angle(\vec{E}, \vec{n})$, $\vec{k}$ is the laser light wave vector, $\vec{B}_{DC}$ is the vector of the static magnetic field, $\vec{E}$ is the directional vector of the semi-major axis of the laser light polarization ellipse, $\vec{n}$ is shown in Figure 1a, and $\varepsilon = |\vec{u}^* \times \vec{u}|$ is the ellipticity parameter with $\vec{u}$ being the normalized Jones vector.

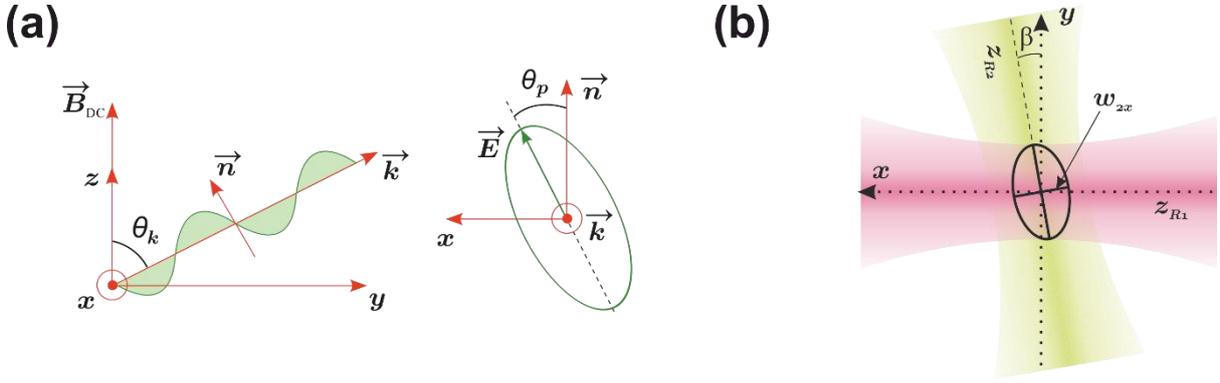

Figure 1. a) Geometry of the ODT beam for polarizability theory. b) Geometry of the cODT in the experiment.

The polarizability coefficients are:

$$\alpha_{sc}(\lambda) = \frac{1}{\sqrt{3(2J+1)}} \alpha_J^{(0)}(\lambda),$$

$$\alpha_{vec}(\lambda) = (-1)^{J+I+F} \sqrt{\frac{2F(2F+1)}{F+1}} \begin{Bmatrix} F & 1 & F \\ J & I & J \end{Bmatrix} \alpha_J^{(1)}(\lambda), \quad (2)$$

$$\alpha_{ten}(\lambda) = (-1)^{J+I+F+1} \sqrt{\frac{2F(2F-1)(2F+1)}{3(F+1)(2F+3)}} \begin{Bmatrix} F & 2 & F \\ J & I & J \end{Bmatrix} \alpha_J^{(2)}(\lambda),$$

where $J$ is an electronic angular-momentum quantum number, $F$ is the total atomic angular momentum quantum number, $I$ is the nuclear spin quantum number; $\alpha_J^{(K)}(\lambda)$, $K \in \{0,1,2\}$ is the coupled polarizability, which can be written as:

$$\alpha_J^{(K)}(\lambda) = \sqrt{2K+1} \sum_{J'} (-1)^{K+1+J+J'} \begin{Bmatrix} 1 & K & 1 \\ J & J' & J \end{Bmatrix} \times$$
$$\left|\langle J' \| \mathbf{d} \| J \rangle\right|^2 \frac{1}{\hbar} \mathrm{Re} \left[ \frac{1}{\Delta_{J'J}^- - i\gamma_{J'}/2} + \frac{(-1)^K}{\Delta_{J'J}^+ - i\gamma_{J'}/2} \right], \quad (3)$$

where $\left|\langle J' \| \mathbf{d} \| J \rangle\right|$ is the reduced dipole matrix element and $\Delta_{J'J}^\pm = 2\pi c/\lambda_{J'J} \pm 2\pi c/\lambda$, $\lambda_{J'J}$ is the wavelength of dipole-allowed transition with the natural linewidth of the excited state $\gamma_{J'}$, and $c$ is the speed of light. The curly brackets denote the Wigner 6-j symbol.

In the case of $\theta_k = 90°$ and linear light polarization, the total polarizability does not depend on the vector part and could be written as:

$$\alpha_{tot}(\lambda) = \alpha_{sc}(\lambda) + \frac{3m_F^2 - F(F+1)}{F(2F-1)} \times \frac{3\cos^2\theta_p - 1}{2} \alpha_{ten}(\lambda). \quad (4)$$

In the other limiting case of $\theta_k = 0°$, the total polarizability does not depend on $\theta_p$:

$$\alpha_{tot}(\lambda) = \alpha_{sc}(\lambda) - \frac{3m_F^2 - F(F+1)}{2F(2F-1)} \alpha_{ten}(\lambda) + \varepsilon \frac{m_F}{2F} \alpha_{vec}(\lambda) \quad (5)$$

Scalar, tensor, and vector atomic polarizabilities at a defined wavelength $\lambda$ can be calculated using second-order perturbation theory [43] by summing the contributions of all known transitions from the National Institute of Standards and Technology database [58].

Atomic polarizability can be measured experimentally using a parabolic approximation of the ODT [51,52] by measuring the parameters of the ODT beam and the frequency of the ODT along one of the axes. In the experiment, we used a crossed configuration of the ODT, which was formed by two horizontal elliptic beams with wavelengths $\lambda_1$ and $\lambda_2$, powers per beam $P_1$ and $P_2$, and waists $w_{1y}$ and $w_{1z}$ for the first beam propogating along the x-axis and waists $w_{2x}$ and $w_{2z}$ for the second beam propogating at an angle $\beta$ to the y-axis (Figure 1b). The frequency $v_{x\beta}$ of the crossed ODT (cODT) can be written in terms of the frequencies $v_{1x}, v_{1y}$ of the first beam and the frequencies $v_{2x}, v_{2y}$ of the second beam:

$$v_{x\beta}^2 = \frac{1}{2}\left(v_{1x}^2 + v_{1y}^2 + v_{2x}^2 + v_{2y}^2\right) +$$

$$+ \frac{1}{2}\sqrt{(v_{1x}^2 - v_{1y}^2)^2 + (v_{2x}^2 - v_{2y}^2)^2 + 2\cos 2\beta (v_{1x}^2 - v_{1y}^2)(v_{2x}^2 - v_{2y}^2)},$$

$$v_{ix} = \frac{1}{2\pi}\sqrt{\frac{2U_i}{mz_{Ri}^2}}, \quad v_{iy} = \frac{1}{2\pi}\sqrt{\frac{4U_i}{mw_{iy}^2}}, \quad z_{Ri}^2 = \frac{2z_{Ryi}^2 z_{Rzi}^2}{z_{Ryi}^2 + z_{Rzi}^2}, \quad z_{Rxi}^2 = \frac{\pi w_{ix}^2}{\lambda_2}, \quad z_{Ryi}^2 = \frac{\pi w_{iy}^2}{\lambda_i}, \quad (6)$$

$$z_{Rzi}^2 = \frac{\pi w_{iz}^2}{\lambda_i}, \quad U_i = \frac{2\pi a_B^3}{c}\alpha_{tot}(\lambda_i)\frac{2P_i}{\pi w_{iy} w_{iz}}, \quad i = 1,2$$

where $a_B$ is a Bohr radius.

In our case, the angle $\beta$ was small (about 2.5°, see Figure 2a) and $v_{1x} \ll v_{2x}$ (see below); therefore, the frequency $v_{x\beta}$ can be written as follows:

$$v_{x\beta} \approx v_{2x}. \tag{7}$$

Total polarizability (in atomic units) can be found as:

$$\alpha_{tot}(\lambda) = \left(\frac{2\pi a_B^3}{c}\frac{2P_2}{\pi w_{2x} w_{2z}}\right)^{-1}\frac{(2\pi v_x(\lambda))^2 w_{2x}^2 m_{Th}}{4}, \tag{8}$$

where $m_{Th}$ is the atomic mass of thulium.

Another way to measure tensor and vector polarizabilities is to measure the differential Stark shift between the hyperfine levels of the states $|F',m'_F\rangle$ and $|F,m_F\rangle$ that arises due to the yODT beam:

$$\delta E_{QSE}(F',m'_F,F,m_F) = \Delta E_{QSE}(F',m'_F) - \Delta E_{QSE}(F,m_F). \tag{9}$$

The polarizability coefficients $\alpha_J^{(0)}(\lambda)$, $\alpha_J^{(1)}(\lambda)$ and $\alpha_J^{(2)}(\lambda)$ take the same values for all sublevels belonging to one atomic level. Therefore, equation (9), using (1) and (2), can be written as follows:

$$\delta E_{QSE} = \delta E_{QSE}^{vec} + \delta E_{QSE}^{ten}$$

$$= -\frac{2\pi a_B^3}{c}\frac{2P_1}{\pi w_{2x} w_{2z}}\left(\alpha_J^{(1)} \varepsilon M^{vec\,F',m'_F}_{F,m_F}\cos\theta_k - \alpha_J^{(2)} M^{ten\,F',m'_F}_{F,m_F}\frac{f(\theta_k,\theta_p,\varepsilon)}{2}\right), \tag{10}$$

where $M^{vec F',m'_F}_{F,m_F}$ and $M^{ten F',m'_F}_{F,m_F}$ are dimensionless coefficients that can be calculated using (1) and (2) and take the values $M^{vec 3,-3}_{4,-4} = -1/48\sqrt{7}$ and $M^{ten 3,-3}_{4,-4} = 1/8\sqrt{35}$ for the $|F=4, m_F=-4\rangle \to |F'=3, m'_F=-3\rangle$ transition.

In the case of the linear light polarization ($\varepsilon = 0$) and $\theta_k = \pi/2$, the dependence of the frequency $\nu_{res}$ of the hyperfine transition $|F=4, m_F=-4\rangle \to |F'=3, m'_F=-3\rangle$ on the angle $\theta_p$ is:

$$\nu_{res} = \nu_{HF} + \nu_{zeeman} + \nu_{QSE}(\theta_p), \quad (11)$$

$$\nu_{QSE}(\theta_p) = \frac{\delta E_{QSE}}{2\pi\hbar} = \frac{1}{2\pi\hbar}\left(-\frac{2\pi a_B^3}{c}\frac{2P_2}{\pi w_{2x} w_{2z}}\right)\alpha_J^{(2)} M^{ten 3,-3}_{4,-4} \frac{1+3\cos(2\theta_p)}{4}, \quad (12)$$

where $\nu_{HF} = 1496.55\,\text{MHz}$ is the hyperfine splitting of the ground state, $\nu_{zeeman} = \frac{1}{2\pi\hbar}(-3g_3 - (-4)g_4)\mu_B B$ is the Zeeman shift, $g_3 = 1.284$ and $g_4 = 0.999$ are the g-factors of the thulium atom, $\mu_B$ is the Bohr magneton, and $B$ is the magnetic field. In our experiment, the magnetic field amplitude was kept constant at a value of 4.464 G, so the Zeeman splitting was approximately 0.9 MHz.

## III. EXPERIMENT

The experiment started with the preparation of an ultracold atomic ensemble of thulium-169 atoms in the cODT formed by two focused laser beams with wavelengths of 1064 nm and 576 nm. Preparation details can be found in previous works [18,19,52,59–63]. A Zeeman slower and 2D optical molasses operating on a strong transition $4f^{13}(^2F^0)6s^2 \to 4f^{12}(^3H_5)5d_{3/2}6s^2$ with a wavelength of 410.6 nm and a natural width of $\Gamma = 2\pi\gamma = 2\pi \cdot 10.5\,\text{MHz}$ provided precooling of the atoms. Then, the atoms were loaded into the magneto-optical trap operating on a weaker transition $4f^{13}(^2F^o)6s^2 \to 4f^{12}(^3H_6)5d_{5/2}6s^2$ with a wavelength of 530.7 nm and a natural width of $\Gamma = 2\pi\gamma = 2\pi \cdot 345.5\,\text{kHz}$. Reducing the magneto-optical trap light intensity to 8 µW, along with increasing the light detuning to $15\Gamma$, polarized the atoms into the lowest magnetic sublevel $|F=4; m_F=-4\rangle$ of the ground state. After cooling the atoms down to $22.5 \pm 2.5$ µK, they were loaded into a transport optical dipole trap (tODT) formed by a linearly polarized laser beam with a waist of 38 μm ($z_{R1} \approx 4.3$ mm) and a wavelength of 1064 nm [52]. The polarized atoms were then moved over a distance of approximately 38 cm into the science chamber using a specific procedure that maintained a constant magnetic field of about 3.8 G to preserve the polarization of the atomic cloud [61].

After transport to the science chamber, there were 6.5 million polarized atoms at a temperature of 55 μK. Subsequent evaporation with the 576 nm laser beam turned on produced around $50-350\times10^3$ atoms (depending on the parameters of the 576 nm laser beam) in the cODT. During evaporation, the power of the tODT beam decreased to the value of 293 mW with $\nu_{1x} = \frac{1}{2\pi}\sqrt{\frac{2U_1}{z_{R1}^2}} = 0.9$ Hz. The magnetic field of 4.464 G remained constant during evaporation.

The horizontal yellow 576 nm laser beam (yODT) with beam waists $w_{2x} = 23\pm1.2$ μm and $w_{2z} = 28.9\pm1.3$ μm had a small angle of 2.5° relative to the y-axis (see Figure 2a). As a laser source, we used a PreciLasers FL-SF-575-1-CW with a central air wavelength of 575.519 nm and a thermal tuning range of ±0.17 nm. The wavelength of the laser beam was monitored with an Angstrom WS-6 wavemeter calibrated using three thulium transitions (410.584 nm, 530.712 nm, and 1139.7951 nm). The horizontality of the yODT was verified using a line laser level and confirmed experimentally with the atoms themselves: after switching off the 1064 nm cODT beam, the center of mass of the ultracold atomic cloud remained at the same position after 4 s of ballistic expansion in the yODT. This effect does not depend on the yODT power (see Figure 2a).

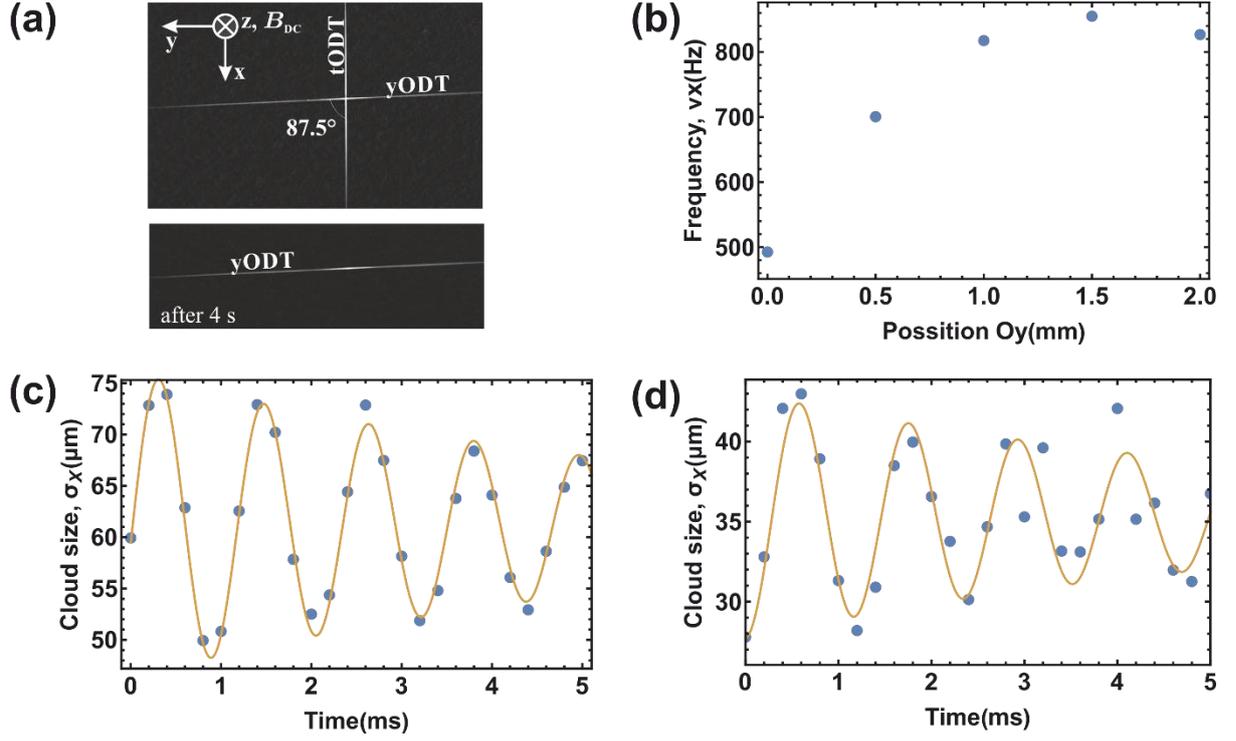

Figure 2. Image of the atoms in the cODT (top). The magnetic field of 4.66 G is oriented along Z. The horizontal tODT beam is oriented along the X-direction, and the horizontal yODT beam makes a small angle of 2.5° with the Y-axis. The bottom image shows the same cloud after 4 s expansion in the yODT. b) Frequency $2\nu_x^{yODT}$ of the yODT versus the y coordinate along the beam. c) Size oscillations of the cloud along the x-axis versus time near the focus (1.5 mm Oy) in the yODT with a fit of the experimental data to an exponentially decaying sine (orange line). The maximum frequency $2\nu_x^{yODT}$ of the yODT at a power of 310 mW is 860 Hz. d) The same oscillations in the cODT after adjusting the intersection of the beams and the focus of the yODT beam (power in the yODT beam is 304 mW). The frequency from the fit is $2\nu_{2x} = 851$ Hz.

After evaporative cooling, size oscillations in the X-direction (see Figure 2b and c) were excited by rapidly doubling the yODT depth, i.e by increasing the yODT beam power with a rise time of ~100 μs. The frequency of the size oscillations corresponds to twice the cODT $\nu_x$ frequency.

The key challenge in the polarizability measurements in the cODT is achieving good overlap between the tODT and the yODT. To accomplish this, we measured the maximum frequency $\nu_x^{yODT}$ in the single yODT (see Figure 2) as a function of the coordinate along the beam as described in [51,52] and adjusted the focus of the yODT beam to intersect the tODT beam. After that, we measured the frequency $v_x$ of the cODT and adjusted the vertical angle of the yODT beam to make $v_x$ equal to the maximum frequency $v_x^{yODT}$, taking into account the beam power. This alignment procedure and the waist measurements were performed at all yODT wavelengths, since after tuning the laser wavelength, we observed a small shift of the waist position along y of about 1 mm.

The dependence of the cODT frequency in the X-direction on the angle $\theta_p$ was measured for different wavelengths of the 576 nm laser. The angle $\theta_p$ was varied using a half-waveplate. At each wavelength of the 576 nm laser, its polarization was verified to be linear in all experiments using a combination of a Thorlabs PM400 power meter with an S121C head and a PBS201 polarizing beamsplitter cube.

The magnetic field of 4.464 G was oriented along the Z-axis and remained constant during all experiments. By knowing the yODT beam parameters (power and waists) and using (8), the total polarizability of thulium atoms was calculated. Figure 3a shows the results for wavelengths of 575.348 nm with a fit using (4) (for the atomic state $|F=4; m_F=-4\rangle$). For the wavelengths 575.445, 575.527, 575.608, and 575.689 nm, the experimental data are shown in Figure 7 (Appendix A, left column). Figure 4a and b show the scalar and tensor polarizabilities obtained from these experiments summarized in Table 1. When the total polarizability is negative, the trap potential becomes repulsive, making frequency measurements impossible. This may create difficulties in fitting the angular dependence of the total polarizability to the experimental data.

A separate RF-assisted trap loss experiment was conducted to directly measure the tensor part of the total polarizability. An RF signal was applied to the cloud in the cODT with an exposure time of 600 ms. The RF frequency was scanned to find the minimum number of atoms remaining in the trap. The typical dependence of the RF-resonance fitted with an inverted Lorentzian function is shown in Figure 3b. The typical resonance width is 1.1 kHz. The magnetic field of 4.464 G was oriented along the Z-axis and remained constant during all experiments. The typical dependence $\Delta v_{QSE}(\theta_p)$ for 575.348 nm is shown in Figure 3c. Using the experimental data (Figure 7, Appendix A, center column) and (12), (2) the tensor part of the polarizability was extracted for the wavelengths of 575.348, 575.445, 575.527, 575.608, and 575.689 nm (Table 1).

To determine the scalar polarizability, we fitted the experimental data (blue lines in Figure 3a and Figure 7, Appendix A, left column) obtained from frequency measurements using expression (4) with the tensor polarizability fixed to the value found from the RF spectroscopy experiments. The resulting scalar polarizabilities are shown in Figure 4a and summarized in Table 1.

The tensor polarizability can also be extracted from the dependence of the total polarizability on the angle $\theta_p$ using equation (12). Despite the absence of data for negative values at some wavelengths, the results for both the scalar and tensor polarizability obtained from RF spectroscopy and frequency experiments are in good agreement (Table 1). Meanwhile, the measured scalar polarizability agrees with the modeled value within the uncertainty, whereas the measured tensor polarizability differs from the theoretical prediction (see Figure 4a and b).

Fit of measured values of polarizability with a Lorentzian function:

$$\alpha(\lambda) = A + \frac{B}{\lambda_0^{-2} - \lambda^{-2}} \tag{13}$$

with the center $\lambda_0 = 576.4287$ nm representing the nearest transition. Equation (13) gives 0 of the total polarizability for the yODT for $\theta_p = 0$ (see Figure 4d) at the wavelength:

$$\lambda_{Tune-out} = 575.646^{+0.016}_{-0.014} \text{ nm.} \tag{14}$$

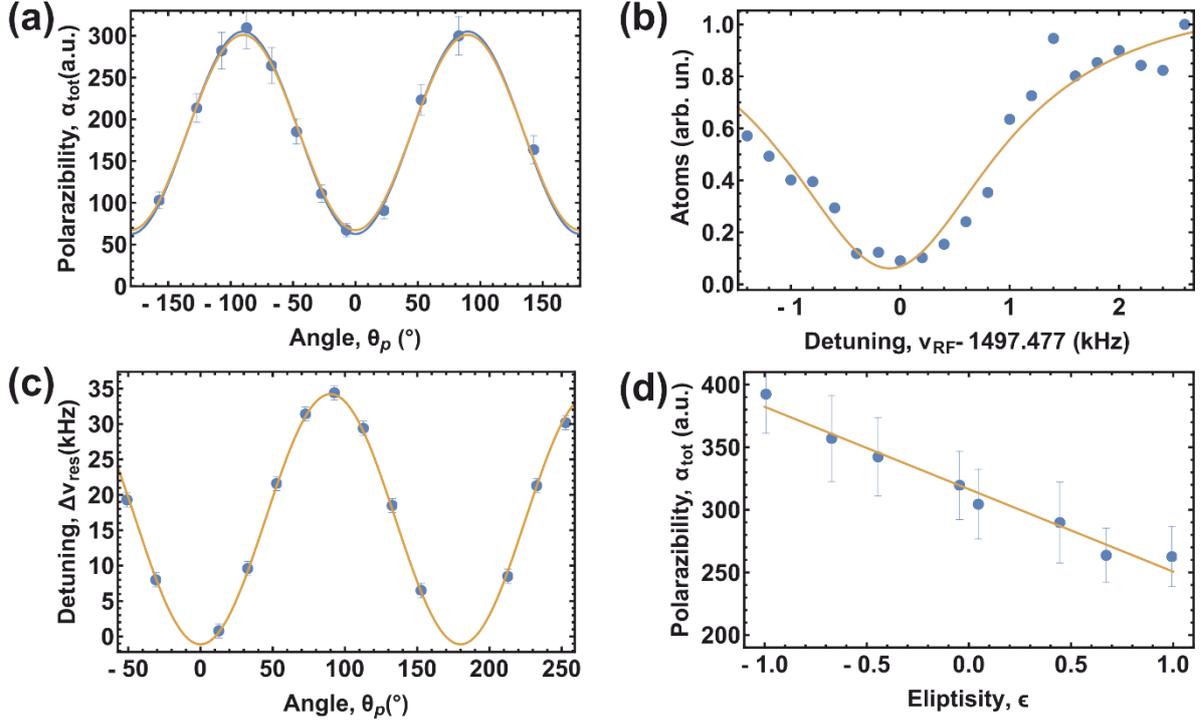

Figure 3. a) Total polarizability of thulium atoms versus the angle $\theta_P$. The orange line is a fit using (4). The dashed black line is a fit to the experimental data with the tensor polarizability fixed to the value obtained from c). b) Number of atoms in the cODT versus the frequency of the RF signal with an exposure time of 600 ms. The solid line is a fit to an inverted Lorentzian function. c) Dependence of the Stark shift between the hyperfine levels of the ground state versus the angle $\theta_P$. The orange line is a fit using (12). d) Total atomic polarizability versus the ellipticity parameter with $\theta_k = 0°$. The orange line is a fit using (5). Error bars include both statistical and systematic uncertainties. The wavelength of yODT is $\lambda_{AIR} = 575.348\,\text{nm}$.

To measure the vector part of the total polarizability, the magnetic field was set to 4.464 G along the yODT direction using three-axis magnetic coils. The coils were calibrated using RF spectroscopy [63–65]. After evaporative cooling in the cODT, size oscillations of the cloud were excited in the X direction in the same manner as before. The oscillation frequency of the atomic ensemble was then measured as a function of the polarization ellipticity of the yODT for various yODT wavelengths. The polarization of the yODT light was controlled by a quarter-wave plate and was verified in all experiments using a Thorlabs PM400 power meter with an S121C head and a PBS201 polarizing beamsplitter cube as follows:

$$\varepsilon = |\vec{u}^* \times \vec{u}| = \frac{2p}{1+p^2},$$

$$p = \sqrt{\frac{P_{MIN}}{P_{MAX}}},$$
(15)

where $P_{MIN}$ and $P_{MAX}$ are the minimum and maximum beam powers at 2 orthogonal orientations of the PBS201. The results for a wavelength of 575.348 nm fitted using (5) assuming the ground state $|F=4; m_F=-4\rangle$ are shown in Figure 3d. The results for the wavelengths of 575.348, 575.445, 575.527, 575.608, and 575.689 nm are presented in Figure 7 (Appendix A, right column). The results for vector polarizability versus wavelength are shown in Figure 4c and summarized in Table 1. Error bars in Figure 4 include statistical and systematic errors. The measured values of the vector polarizability are in good agreement with the model within uncertainties.

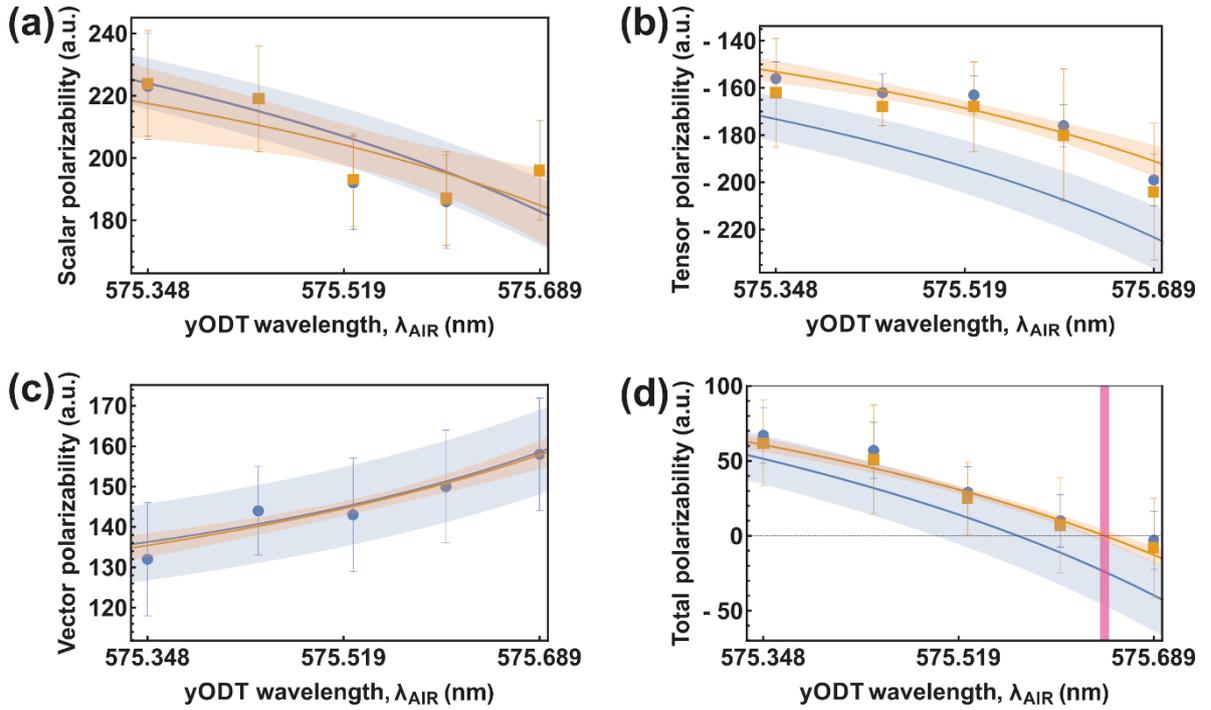

Figure 4. Scalar (a), tensor (b), and vector (c) polarizabilities versus wavelength. The dots represent measured values. Orange squares correspond to RF spectroscopy measurements, and blue dots to ODT frequency measurements. The blue lines show the second-order perturbation theory model, taking into account the uncertainty in the transition probabilities [66], indicated by the blue shaded area. The orange lines are Lorentzian fits with a center at $\lambda_0 = 576.4287$ nm. d) Total polarizability of thulium atoms in the state $F=4, m_F=-4$ versus

wavelength in the configuration $\varepsilon = 0$, $\theta_P = 0°$, $\theta_k = 90°$. The pink area indicates the range where the total polarizability becomes zero, as measured in the experiment (Figure 5).

To directly observe the negative total polarizability of the yODT in the configuration $\theta_p = 0°, \theta_k = 90°$, the number of atoms in the cODT was measured as a function of the yODT wavelength. As clearly seen in Figure 5, at a wavelength of 575.650 nm, atoms disappear from the trap, while remaining in the tODT. This absence of atoms indicates a negative total polarizability in the cODT (see Figure 5). The gap corresponds to the repulsive potential of the yODT. The wavelength of 575.642 nm corresponds to the borderline case, where the cODT is still visible on top of the tODT, but a small peak corresponding to a small fraction of atoms remains in the cODT, demonstrating nearly zero but still positive polarizability. The yODT polarizability is clearly negative at 575.65 nm and positive at 574.642 nm, which brackets the measured value of $575.646^{+0.016}_{-0.014}$ nm (14) for 0 polarizability and in fact makes it possible to narrow the value for turn-out wavelength to $575.646^{+0.004}_{-0.004}$ nm. The model prediction for 0 polarizability is 575.568±0.07 nm (see Figure 4d), which is consistent with the measurements.

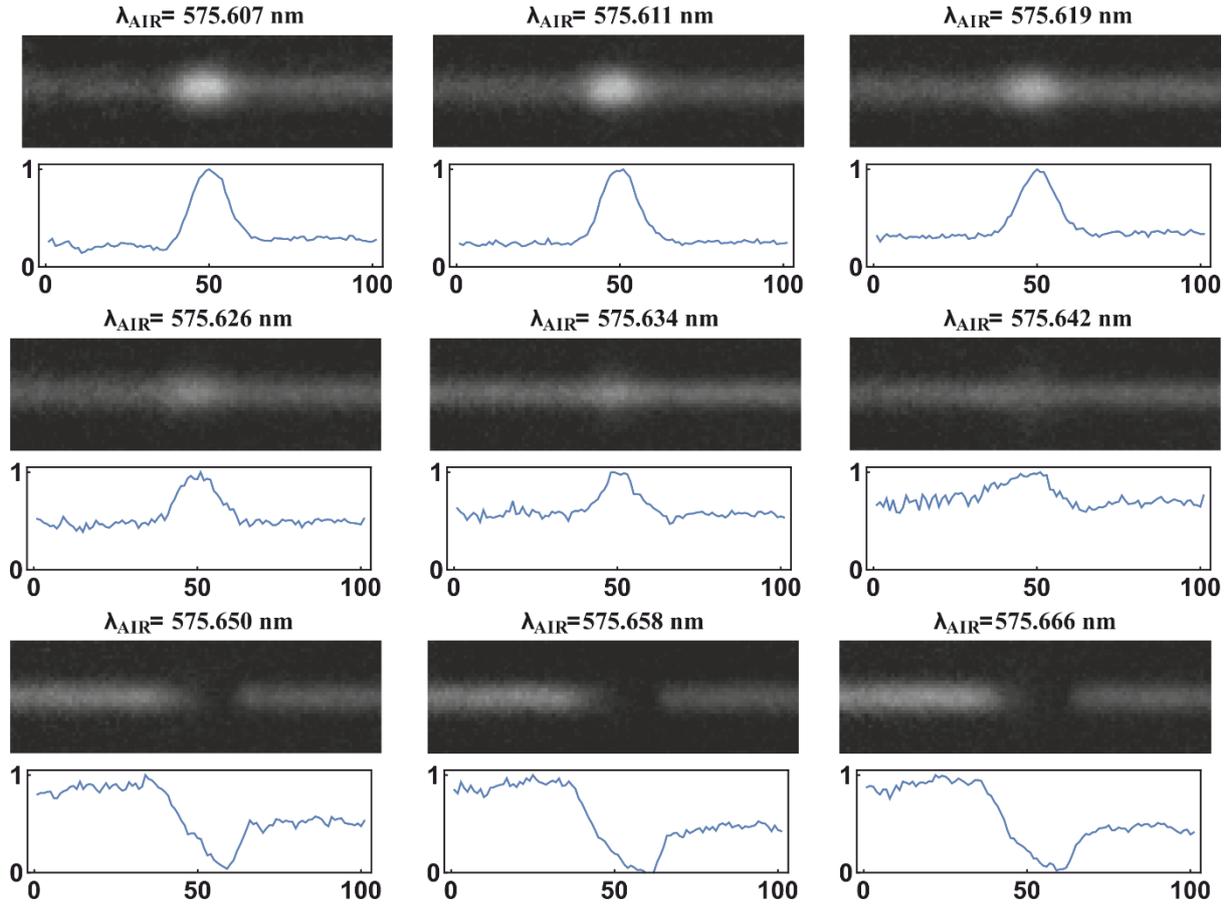

Figure 5. Images of the atomic clouds of thulium atoms (top images) with the normalized atomic distributions along the horizontal axis (bottom images) in the cODT for different wavelengths of the yODT ($\theta_P = 0°$) beam.

To ensure that, in the wavelength range of 575.348–575.689 nm, there is no significant heating of the atoms due to photon scattering (i.e the imaginary part of the polarizability), the yODT in the configuration $\theta_p = 90°$, $\theta_k = 90°$ was implemented. The evaporation sequence in the cODT was performed, successfully reaching the Bose-Einstein condensation (BEC) of thulium atoms at each wavelength. As an example, Figure 6a shows an image of the BEC of thulium atoms after 16 ms of ballistic expansion from the cODT (with the yODT wavelength set to 575.689 nm). Figure 6b presents a bimodal fit [67] of the atomic distribution shown in Figure 6a, with $16.5×10^3$ thulium atoms in the BEC.

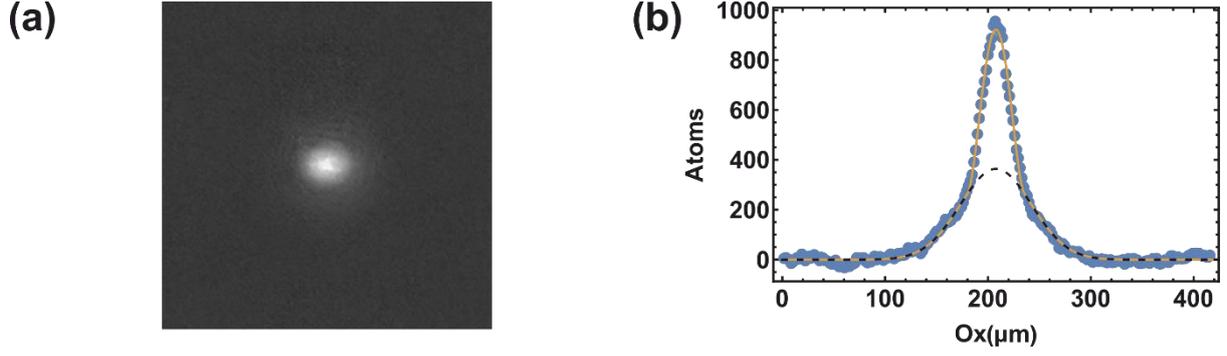

Figure 6. a) Image of the BEC of thulium atoms after 16 ms of ballistic expansion from the cODT with yODT parameters $\lambda = 575.689$ nm, $\theta_P = 90°$. b) The orange line represents a fit of the atomic distribution along the vertical axis with a bimodal distribution; the dashed shows a Gaussian fit to the thermal cloud. The number of atoms in the BEC is approximately $16.5 \times 10^3$.

Table 1. Polarizability of Tm-169 at different wavelengths.

| Polarizability, Air wavelength, nm | $\alpha_{scal}$, (a.u.) | | | $\alpha_{tens}$, (a.u.) | | | $\alpha_{vec}$, (a.u.) | |
|---|---|---|---|---|---|---|---|---|
| | Theory | Exp. freq. | Exp. RF spec. and freq. | Theory | Exp. freq. | Exp. RF spec. | Theory | Exp. freq. |
| 575.348 | 223 | 224±17 | 223±17 | -175 | -162±23 | -156±7 | 137 | 132±14 |
| 575.444 | 213 | 219±17 | 219±17 | -186 | -168±32 | -162±8 | 142 | 144±11 |
| 575.526 | 204 | 193±15 | 192±15 | -197 | -168±19 | -163±8 | 147 | 143±14 |
| 575.608 | 193 | 187±15 | 186±15 | -210 | -180±28 | -176±9 | 153 | 150±14 |
| 575.689 | 180 | 196±16 | 196±16 | -227 | -204±29 | -199±11 | 160 | 158±14 |

## IV. CONCLUSION

The scalar, tensor, and vector polarizabilities of the thulium atom were calculated near a wavelength of 576 nm based on 59 known atomic transitions of thulium. The total polarizability was then measured in the range of 575.348–575.689 nm by measuring the trap frequencies in the ODT formed by the intersection of the 576 nm and 1064 nm laser beams. The tensor polarizability was measured using RF spectroscopy, making it possible to extract the scalar part of the total polarizability. The obtained values of the scalar and vector polarizabilities are in very good agreement with the model, while the tensor polarizability shows a slight deviation. Nevertheless, a direct fit of the angular dependence of the total frequency is consistent with the RF spectroscopy results. The zero value was confirmed to be around the predicted wavelength of $575.646^{+0.016}_{-0.014}$ nm. Moreover, the zero of the polarizability and its negative region were directly observed by monitoring the number of atoms in the trap as a function of the 576 nm trap wavelength in the configuration $\theta_p = 0°$, $\theta_k = 90°$ which made it possible to refine the value to $575.646^{+0.004}_{-0.004}$. The BEC of thulium atoms was achieved at each wavelength, indicating that heating due to photon scattering (imaginary part of the polarizability) is negligible.

## V. ACKNOWLEDGEMENTS

This work was supported by Rosatom in the framework of the Roadmap for Quantum computing (Contract No. 868/1759-D dated October 3, 2025).

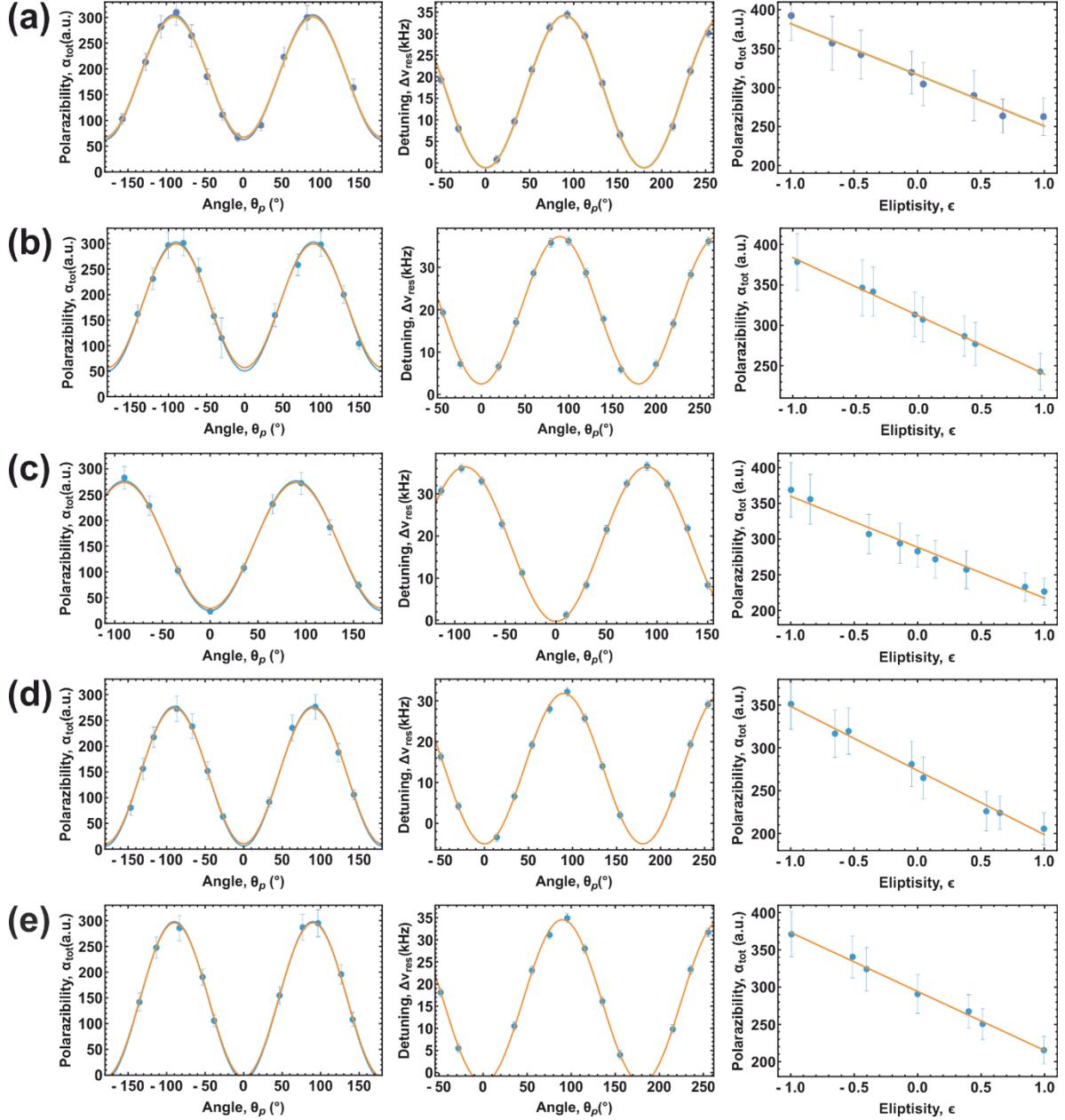

Figure 7. Experimental data for scalar, tensor, and vector polarizabilities in experiments with RF spectroscopy (center column) and trap frequency measurements (left and right columns) for different wavelengths of the yODT beam. Points are experimental data. Orange lines are fits by (12) for RF spectroscopy data, (4) for scalar and tensor polarizabilities, and (5) for vector polarizability. Blue lines are fits by (4) for fixed tensor polarizability that was found from the RF spectroscopy experiment. a) 575.348 nm. b) 575.445 nm. c) 575.526 nm. d) 575.608 nm. e) 575.689 nm.